\documentclass{article}
\pdfoutput=1

\usepackage{amstext,amsmath,amssymb,amsthm,tikz,tikz-cd}
\usepackage{array,color,stmaryrd}
\usepackage{hyperref}

\newtheorem{fact}{Fact}
\newtheorem{proposition}{Proposition}
\newtheorem{lemma}{Lemma}
\newtheorem{theorem}{Theorem}
\newtheorem{corollary}{Corollary}

\newcommand{\states}{\mathcal{S}}

\newcommand{\compl}[1]{\overline{#1}}
\newcommand{\acompl}{{\sim}}
\newcommand{\ocompl}[1]{{#1}^{\bot}}
\newcommand{\OML}{\mathfrak{OML}}

\newcommand{\posop}[1]{\langle#1\rangle}
\newcommand{\necop}[1]{{[}#1{]}}
\newcommand{\BAO}{\mathfrak{BAO}}

\newcommand{\dedBQ}{\vdash_{\mathrm{BQ}}}

\title{Quantum Logic as Classical Logic\thanks{I dedicate this paper to my professor Jacques Zahnd.}}
\author{Simon Kramer\\[\jot]
		SK-R\&D Ltd liab.\ Co\\
		\texttt{simon.kramer@a3.epfl.ch}}

\begin{document}
\maketitle

\begin{abstract}
	We propose a semantic representation of the standard quantum logic QL within 
		a classical, normal modal logic, and this via
			a \emph{lattice-embedding} of 
				orthomodular lattices into 
				Boolean algebras with one modal operator.
	Thus our classical logic is a \emph{completion} of the quantum logic QL.
	In other words, 
		we refute Birkhoff and von Neumann's classic thesis that 
			the logic (the formal character) of Quantum Mechanics would be non-classical as well as 
		Putnam's thesis that quantum logic (of his kind) would be the correct logic for propositional inference in general.
	The propositional logic of Quantum Mechanics is modal but classical, and
		the correct logic for propositional inference need not have an extroverted quantum character.
	One normal necessity modality $\Box$ 
	suffices 
		to capture the subjectivity of observation in quantum experiments, and this 
			thanks to its failure to distribute over classical disjunction.
	The key to our result is the translation of 
		\emph{quantum negation as classical negation of observability.}
	
\end{abstract}

\section{Introduction}
The idea for this paper originates in an observation that 
	I made in the plenum of 
		a talk delivered by Newton da Costa at 
			the 4th World Congress and School on Universal Logic in Rio de Janeiro in 2013.
The observation is about a presentation of a quantum experiment that 
	has been put forward  
		as a counter-example to the suitability of classical logic for reasoning about quantum phenomena and 
		as a motivation for alternative logics such as quantum logics, over and over again.
The presentation usually involves three statements, say $P$, $Q$, and $R$,  
	each one being about an elementary quantum phenomenon produced by the experiment, 
		but such that 
			\begin{equation}
				\underbrace{\text{$\text{$\underbrace{\text{$P$ is observed to be true}}_{\text{$\Box P$ is true}}$ and 
						$\underbrace{\text{($Q$ or $R$) is observed to be true}}_{\text{$\Box(Q\lor R)$ is true}}$}$}}_{%
							\text{$\Box P\land\Box(Q\lor R)$ is true}}.
			\end{equation}
Notice that the observation of the truth of a disjunction does not imply 
	the observation of the truth of one of its disjuncts.
That is,
	$\Box(Q\lor R)\rightarrow(\Box Q\lor\Box R)$ is \emph{not} a valid principle. 
This is an essential uncertainty.\footnote{This is also an important difference to 
	Intuitionistic Logic (and others), where the validity of a disjunction does imply 
		the validity of at least one disjunct (cf.\ Disjunction Property).}
(On the other hand,
	the converse $(\Box Q\lor\Box R)\rightarrow\Box(Q\lor R)$ \emph{is} a valid principle.)
Hence, and in fact,  
	\begin{equation}
		\underbrace{\text{neither ($P$ and $Q$) nor ($P$ and $R$) is observed to be true}.}_{\text{%
		$\neg\Box(P\land Q)\land\neg\Box(P\land R)$ is true\quad 
		(or: $\neg(\Box(P\land Q)\lor\Box(P\land R))$ is true)}}
	\end{equation}
The quantum-mechanical details need not concern us here.
However,
	what does need concern us here is that 
		the presentation of the experiment concludes that 
			$$\text{($P$ and ($Q$ or $R$)) is true but 
				not (($P$ and $Q$) or ($P$ and $R$)).}$$
That is,
	\begin{equation}
		\text{``($P$ and ($Q$ or $R$))'' and ``(($P$ and $Q$) or ($P$ and $R$))'' are not equivalent.}
	\end{equation}
Apparently, 
	the distributivity of  
		classical conjunction and disjunction fails!
Whence arises the motivation for special \emph{quantum} conjunction and disjunction.

Now, my observation is that 
	the experiment---though classic---is not well presented, that is, 
		the formalisation of the experimental observations is unfortunate.
The point is that 
	the fact of \emph{observing} the fact $P$ and the fact ($Q$ or $R$) 
		should be made explicit in the formalisation too, for example as 
			$\Box P\land\Box(Q\lor R)$.
Hence in the experiment,
	$\Box P\land\Box(Q\lor R)$ is true (1) but not $\Box(P\land Q)\lor\Box(P\land R)$ (2).
That is,
	$(\Box P\land\Box(Q\lor R))\rightarrow(\Box(P\land Q)\lor\Box(P\land R))$ is false.
On the other hand,
	the converse $(\Box(P\land Q)\lor\Box(P\land R))\rightarrow(\Box P\land\Box(Q\lor R))$ is true, because:
		$$\begin{array}{@{}r@{\;}c@{\;}l@{}}
			 (\Box(P\land Q)\lor\Box(P\land R))&\rightarrow& \Box((P\land Q)\lor(P\land R)) \text{ is true}\\[2\jot]
			 	&\leftrightarrow& \Box(P\land (Q\lor R)) \text{ is true}\\[2\jot]
			 	&\leftrightarrow& (\Box P\land\Box(Q\lor R)) \text{ is true}
		\end{array}$$		
(As noticed above,
	$\Box$ distributes over $\land$ in both directions, but over $\lor$ only in one direction.)
Thus, and in close correspondence with (3), 
	\begin{equation}
		\text{$(\Box P\land\Box(Q\lor R))\leftrightarrow(\Box(P\land Q)\lor\Box(P\land R))$ is false.}
	\end{equation}
Hence,
	if we \emph{\textbf{make explicit the fact of observing facts}} (for example by means of a modal operator $\Box$) 
	then we do not need to introduce the special-purpose formalism of Quantum Logic with 
		special and possibly counter-intuitive quantum operators 
			to account for quantum phenomena (due to the apparent failure of classical conjunction to distribute over classical disjunction), but 
				can get by with intuitive classical (Boolean) logic at the small price of 
					adding a single, classical modal operator $\Box$\,. 
This operator is characterised by the following 
	two axioms K (normality) and BQ plus the single deduction rule N (normality):
	\begin{enumerate}
		\item $\Box(A\rightarrow B)\rightarrow(\Box A\rightarrow\Box B)$\qquad(Kripke's law, K)
		\item $\Box\Diamond A\leftrightarrow A$, where $\Diamond := \neg\Box\neg$\qquad(BQ)
		\item from $A$ infer $\Box A$\qquad(necessitation rule, N)
	\end{enumerate}
BQ has the encouraging meaning that quantum truth is equivalent to
	\begin{itemize}
		\item observing the possibility of that truth, or, equivalently, 
		\item the possibility of observing that truth.
	\end{itemize}
(Consider that 
	$\Box\Diamond A\leftrightarrow A$ is true if and only if 
	$\neg\Box\Diamond A\leftrightarrow\neg A$ is true if and only if 
	$\Diamond\neg\Diamond A\leftrightarrow\neg A$ is true if and only if 
	$\Diamond\Box\neg A\leftrightarrow\neg A$ is true if and only if 
	$\Diamond\Box A\leftrightarrow A$ is true.)
The laws of Boolean logic plus 
the three modal laws K, BQ, and N, which are embodied in our classical modal logic BQ, 
suffice 
	to model the logical essence of Quantum Mechanics as captured by 
		the standard quantum logic QL \cite[Page~32]{HBQuantumLogic}.
To appreciate the smallness of the price of understanding BQ, 
	consider the price of understanding QL.
This standard quantum logic is characterised by 
	the following 12 axioms plus one deduction rule \cite[Page~32]{HBQuantumLogic}:
\begin{enumerate}
	\item $( A \equiv B )\rightarrow_{0}(( B \equiv C )\rightarrow_{0}( A \equiv C ))$
	\item $( A \equiv B )\rightarrow_{0}(\acompl A \equiv\acompl B )$
	\item $( A \equiv B )\rightarrow_{0}(( A \curlywedge C )\equiv( B \curlywedge C ))$
	\item $( A \curlywedge B )\equiv( B \curlywedge A )$
	\item $( A \curlywedge ( B \curlywedge C ))\equiv(( A \curlywedge B )\curlywedge C )$
	\item $( A \curlywedge ( A \curlyvee B ))\equiv A $
	\item $(\acompl  A \curlywedge A )\equiv((\acompl A \curlywedge A )\curlywedge B )$
	\item $A \equiv\acompl\acompl A $
	\item $\acompl( A \curlyvee B )\equiv(\acompl A \curlywedge\acompl B )$\qquad(De-Morgan law)
	\item $( A \equiv B )\equiv( B \equiv A )$
	\item $( A \equiv B )\rightarrow_{0}( A \rightarrow_{0} B )$
	\item $( A \rightarrow_{0} B )\rightarrow_{3}( A \rightarrow_{3}( A \rightarrow_{3} B ))$
	\item from $A$ and $A \rightarrow_{3} B$ infer $B$
\end{enumerate}
with the three abbreviations: 
	$$\begin{array}[t]{@{}r@{\ \ }c@{\ \ }l@{}}
				 A \rightarrow_{0} B &:=&\acompl A \curlyvee B\\[\jot]
				 A \rightarrow_{3} B &:=&(\acompl A \curlywedge B )\curlyvee(\acompl A \curlywedge\acompl B )\curlyvee( A \curlywedge(\acompl  A \curlyvee B ))\\[\jot]
				 A \equiv B &:=&( A \curlywedge B )\curlyvee(\acompl A \curlywedge\acompl B )
		\end{array}$$
and $\curlywedge$, $\curlyvee$, and $\acompl$ symbolising 
	quantum conjunction, quantum disjunction, and quantum negation, respectively 
		(my choice, to distinguish the quantum operators clearly from their classical counterparts).
Observe that 
	this axiom system for QL employs 
		two notions of implication, symbolised as $\rightarrow_{0}$ and $\rightarrow_{3}$, as well as 
		one notion of equivalence, symbolised as $\equiv$\,, but that 
			is defined in terms of neither notion of implication!
Proving the adequacy of this axiom system for 
	the standard quantum structures of orthomodular lattices  
	must have been a real \emph{tour de force,} which 
	must be appreciated as such.
Fortunately,
	we do not need to understand nor use this axiom system in order to 
		understand the logical essence of Quantum Mechanics.
All we need to understand is that 
	QL and its corresponding, standard quantum-structure semantics of 
		orthomodular lattices satisfy the De-Morgan law, and thus 
			quantum disjunction is definable in terms of quantum negation and quantum conjunction.
Hence in essence, 
	the culprit for the failure of quantum conjunction 
		to distribute over quantum disjunction boils down to quantum negation!
We thus must find a translation of quantum negation in terms of 
	the modal operator $\Box$ and Boolean operators.
The translation that we have found and shall now present and explicate is to 
	$$\boxed{\text{translate quantum negation $\acompl$ as $\neg\Box$\,.}}$$
That is,
	we translate \emph{quantum negation as classical negation of observability.}
Recall from classical normal modal logic that 
	$\neg\Box$ (``not necessarily'') is the same as $\Diamond\neg$ (``possibly not'').
Hence, 
	the classical negation of observability is 
		classically equivalent to the possibility of observing classical negation.
Thus, 
	we can also view \emph{quantum negation as 
		the possibility of observing classical negation.}

\section{The technical details of the translation}
We shall carry out our translation from QL into BQ semantically, by
	producing a \emph{lattice-embedding} from 
		the standard orthomodular-lattice (OML-)model of QL into 
		the standard Boolean-algebra-with-one-operator (BAO-)model of BQ \cite{AlgebrasAndCoalgebras}.
So, not only logicians but also mathematicians will be able to appreciate our result.
For simplicity,
	we will reuse some of the symbols 
		for the (syntactic) operators of QL from the introduction 
		for their corresponding (semantic) operators of the OML-model of QL, and
	only use different symbols for 
		the (syntactic) operators of BQ and 
		their corresponding (semantic) operators of our BAO.
Both algebraic models involve a set of subsets of a set of states as carrier together with 
	algebraic operations on this carrier set.
So, 
	let 
		$\states$ designate our \emph{state space,} that is, 
			the set of all possible worlds, points, or states, 
		$\mathrm{H}(\states)$ the set of subsets of $\states$ that 
			is algebraically closed under the OML-operations for 
			the carrier of the OML-model, and 
		$\mathrm{P}(\states)$ the powerset of $\states$ for 
			the carrier of our BAO-model.

Then, translate 
	the \emph{\textbf{orthomodular}} (and thus \emph{De Morgan}) \emph{\textbf{lattice}} \cite{HBQuantumLogic} 
		$$\OML\ :=\ 
			\langle\,\mathrm{H}(\states),0,\curlywedge, \curlyvee,1,\ocompl{\cdot},\preccurlyeq\,\rangle$$ 
	on $\states$ to a corresponding (inclusion-ordered, complete) \emph{\textbf{Boolean} (powerset) \textbf{algebra} (lattice)}  
		$$\BAO\ :=\ 
			\langle\,\mathrm{P}(\states), \emptyset, \cap, \cup, \states, \compl{\,\cdot\,},\posop{R},\subseteq\,\rangle$$ 
	 \emph{\textbf{with one operator}} 
		$\posop{R}:2^\states\to2^\states$  
		$$\posop{R}(S)\ :=\ 
				\{\,s\in\states\mid\text{there is $s'\in\states$ such that $s\mathrel{R}s'$ and $s'\in S$}\,\}$$ for an---only for now---arbitrary (see the end of this section) accessibility relation\footnote{%
					Accessibility relations are at the heart of Kripke-semantics for modal logics \cite{ModalLogicSemanticPerspective}.} $R\subseteq\states\times\states$ \cite{AlgebrasAndCoalgebras}, that is, 
						a binary relation on $\states$ of no particularity, and with 
	dual operator 
			$\necop{R}:2^\states\to2^\states$ \cite[Definition~3.8.2]{PracticalFoundationsOfMathematics}
			$$\necop{R}(S)\ :=\  
				\{\,s\in\states\mid\text{for all $s'\in\states$, if $s\mathrel{R}s'$ then $s'\in S$}\,\}$$  
		by means of 
		an \emph{injective} mapping $\rho:\OML\rightarrowtail\BAO$ such that 
	$$\begin{array}{@{}r@{\ \ }c@{\ \ }l@{}}
	\rho(\ocompl{H}) &=& \acompl\rho(H)\qquad\text{($\acompl$-homomorphism)}\\[\jot]
	\rho(H\curlywedge H') &=& \rho(H)\cap\rho(H')\qquad\text{(meet homomorphism)}
\end{array}$$	
	where $\acompl:=\posop{R}\circ\compl{\,\cdot\,}$. 
As usual, $\circ$ designates function composition.
Note that $\rho$ is induced (and thus exists) by 
	its syntactic counterpart mentioned in the introduction 
		(translate $\acompl$ as $\neg\Box$, or, equivalently,  as $\Diamond\neg$), and this via
			the standard Lindenbaum-Tarski-algebra construction \cite{DaveyPriestley} 
				on the language of QL and on the language of BQ.
(See also Corollary~\ref{corollary:reduction} for this translation.)

The operator $\necop{R}$ is the semantic analog of the modality $\Box$, and 
	is related to $R$ as asserted by the following fact.
\begin{fact}[{\cite[Exercise~3.65]{PracticalFoundationsOfMathematics}}] 
$$\text{$s\mathrel{R}s'$ if and only if 
	for all $S\subseteq\states$, $s\in\necop{R}(S)$ implies $s'\in S$}$$
\end{fact}
\noindent
As opposed to the operator $\necop{R}$, 
	its dual $\posop{R}$ does distribute over set union $\cup$, 
		as asserted by the following well-known, but to our development crucial fact.
\begin{fact}[Property of $\posop{R}$]\label{fact:PropertiesOfPosOp} 
	$$\posop{R}(S\cup S')=\posop{R}(S)\cup\posop{R}(S')$$
\end{fact}
\noindent
Now, recall the following laws of orthomodular lattices: 
\begin{itemize}
	\item De Morgan: 
			$$H\curlyvee H' = \ocompl{(\ocompl{H}\curlywedge\ocompl{H'})}$$
	\item orthocomplementarity: 
		\begin{itemize}
			\item involution: $\ocompl{\ocompl{H}}=H$
			\item disjointness: $H\curlywedge\ocompl{H}=0$
			\item exhaustiveness: $H\curlyvee\ocompl{H}=1$
			\item antitonicity: $H\preccurlyeq H'\Rightarrow\ocompl{H'}\preccurlyeq\ocompl{H}$
		\end{itemize}
	\item orthomodularity (OM): 
			$$\begin{array}{@{}r@{\ \ }c@{\ \ }l@{\ }r@{\ }l@{}}
				H\preccurlyeq H' &\text{:iff}& H&=&H\curlywedge H'\\[\jot]
								 &\Leftrightarrow& H'&=&H\curlyvee H'\\[\jot]
								 &\Rightarrow& H'&=&H\curlyvee(H'\curlywedge\ocompl{H})\qquad\text{(OM)}
			\end{array}$$
\end{itemize}

\begin{proposition}\label{proposition:Completion}
	The complete lattice $\BAO$ is a \emph{completion} \cite[Definition~7.36]{DaveyPriestley} of 
		the lattice (and thus partially ordered set) $\OML$ via 
			the \emph{\textbf{order}-embedding} $\rho$, that is, 
	$$\text{for all $H,H'\in\mathrm{H}(\states)$,
		$H\preccurlyeq H'$ if and only if $\rho(H)\subseteq\rho(H')$\,.}$$
\end{proposition}
\begin{proof}
	Let $H,H'\in\mathrm{H}(\states)$.
	Suppose that $H\preccurlyeq H'$.
	By definition of $\preccurlyeq$, 
		$H=H\curlywedge H'$.
	Further suppose that $s\in\rho(H)$.
	Thus $s\in\rho(H\curlywedge H')$.
	Hence $s\in\rho(H)\cap\rho(H')$ by the meet-homomorphism property of $\rho$.
	Thus $s\in\rho(H')$.
	Conversely, suppose that $\rho(H)\subseteq\rho(H')$.
	Thus $\rho(H)=\rho(H)\cap\rho(H')$.
	Hence $\rho(H)=\rho(H\curlywedge H')$ by the meet-homomorphism property of $\rho$.
	Hence $H=H\curlywedge H'$ by the injectivity of $\rho$.
	Thus $H\preccurlyeq H'$.
\end{proof}
\noindent
Notice that 
	we have not used any property of $\acompl$ in the proof of Proposition~\ref{proposition:Completion}.
The interesting properties of $\acompl$ are enumerated in the next proposition, which
	asserts that $\acompl$ can mimic the orthocomplement $\ocompl{}$ of $\OML$ in our $\BAO$ via $\rho$.
The essential ones are the first and the second 
	(see the proofs of the others).
\begin{proposition}[Properties of $\acompl{}$]\label{proposition:PropertiesOfSim}\ 
	\begin{enumerate}
		\item \colorbox[gray]{0.75}{$\acompl\acompl\rho(H)=\rho(H)$\qquad(involutive interaction with itself)}
		\item \colorbox[gray]{0.75}{$\acompl(S\cap S')=\acompl S\cup\acompl S'$\qquad(De-Morgan interaction with meet and join)}
		
				(thus $\acompl(\rho(H)\cap\rho(H'))=\acompl\rho(H)\cup\acompl\rho(H')$)
		\item $\acompl(\rho(H)\cup\rho(H'))=\acompl\rho(H)\cap\acompl\rho(H')$
		\item $\rho(H)\cap\acompl{\rho(H)}=\rho(0)$
		\item $\rho(H)\cup\acompl{\rho(H)}=\rho(1)$
		\item $(H\preccurlyeq H'\text{ or }\rho(H)\subseteq\rho(H'))$ implies 
				\begin{enumerate}
					\item $\acompl\rho(H')\subseteq\acompl\rho(H)$ and 
					\item $\rho(H')=\rho(H)\cup(\rho(H')\cap\acompl{\rho(H)})$
				\end{enumerate}
		\item $\acompl\rho(0)=\rho(1)$
		\item $\acompl\rho(1)=\rho(0)$
	\end{enumerate}
\end{proposition}
\begin{proof}
	For (1),   
		let $H\in\mathrm{H}(\states)$ and 
		recall that $\ocompl{\ocompl{H}}=H$.
	Thus $\rho(\ocompl{\ocompl{H}})=\rho(H)$.
	Hence $\acompl{\acompl{\rho(H)}}=\rho(H)$.
	For (2), 
		let $S,S'\in\mathrm{P}(\states)$ and 
		consider: 
	$$\begin{array}{@{}r@{\ \ }c@{\ \ }l@{\qquad}l@{}}
		\acompl(S\cap S')	&=& \acompl\left(\,\compl{\compl{S}\cup\compl{S'}}\,\right) & \text{([Boolean] De Morgan)}\\[2\jot]
							&=& (\posop{R}\circ\compl{\,\cdot\,}\circ\compl{\,\cdot\,}\,)\left(\,\compl{S}\cup\compl{S'}\,\right) & \text{(definition)}\\[\jot]
							&=& \posop{R}\left(\,\compl{S}\cup\compl{S'}\,\right) & \text{([Boolean] involution)}\\[\jot]
							&=& \posop{R}\left(\,\compl{S}\,\right)\cup\posop{R}\left(\,\compl{S'}\,\right) & \text{(Fact~\ref{fact:PropertiesOfPosOp})}\\[\jot]
							&=& (\posop{R}\circ\compl{\,\cdot\,}\,)(S)\cup(\posop{R}\circ\compl{\,\cdot\,}\,)(S') & \text{(definition)}\\[\jot]
							&=& \acompl S\cup\acompl S' & \text{(definition)}
	\end{array}$$
	For (3), 
		let $H,H'\in\mathrm{H}(\states)$ and 
		consider: 
	$$\begin{array}{@{}r@{\ \ }c@{\ \ }l@{\qquad}l@{}}
		\acompl(\rho(H)\cup\rho(H'))	&=& \acompl(\acompl(\acompl\rho(H))\cup\acompl(\acompl\rho(H'))) & \text{(1)}\\[\jot]
							&=& \acompl\acompl(\acompl\rho(H)\cap\acompl\rho(H')) & \text{(2)}\\[\jot]
							&=& \acompl\rho(H)\cap\acompl\rho(H') & \text{(1)}
	\end{array}$$
	For (4) and (5), let $H\in\mathrm{H}(\states)$.
	For (4), recall that $H\curlywedge\ocompl{H}=0$.
	Thus $\rho(H\curlywedge\ocompl{H})=\rho(0)$.
	Hence,  
		$\rho(H)\cap\rho(\ocompl{H})=\rho(0)$ and then
		$\rho(H)\cap\acompl{\rho(H)}=\rho(0)$.
	For (5), recall that $H\curlyvee\ocompl{H}=1$.
	Hence:
	$$\begin{array}{@{}r@{\ \ }c@{\ \ }l@{\qquad}l@{}}
		\rho(1) &=& \rho(H\curlyvee\ocompl{H}) & \\[2\jot]
				&=& \rho\left(\ocompl{\left(\ocompl{H}\curlywedge\ocompl{\ocompl{H}}\right)}\right) & \text{(definition)}\\[4\jot]
				&=& \rho\left(\ocompl{(\ocompl{H}\curlywedge H)}\right) & \text{(involution)}\\[2\jot]
				&=& \acompl{\rho(\ocompl{H}\curlywedge H)} & \text{($\acompl$-homomorphism)}\\[\jot]
				&=& \acompl{(\rho(\ocompl{H})\cap\rho(H))} & \text{(meet-homomorphism)}\\[\jot]
				&=& \acompl{\rho(\ocompl{H})}\cup\acompl{\rho(H)} & \text{(2)}\\[\jot]
				&=& \acompl{\acompl{\rho(H)}}\cup\acompl{\rho(H)} & \text{($\acompl$-homomorphism)}\\[\jot]
				&=& \rho(H)\cup\acompl{\rho(H)} & \text{(1)}
	\end{array}$$
	For (6.a) and (6.b),
		let $H,H'\in\mathrm{H}(\states)$ and 
		suppose that $H\preccurlyeq H'\text{ or }\rho(H)\subseteq\rho(H')$.
	For (6.a), 
		first suppose that $H\preccurlyeq H'$.
	Thus $\ocompl{H'}\preccurlyeq\ocompl{H}$ by antitonicity.
	Hence $\rho(\ocompl{H'})\subseteq\rho(\ocompl{H})$ by Proposition~\ref{proposition:Completion}.
	Hence $\acompl{\rho(H')}\subseteq\acompl{\rho(H)}$ by $\acompl$-homomorphism.
	Now suppose for (6.a) that $\rho(H)\subseteq\rho(H')$.
	Hence $H\preccurlyeq H'$ by Proposition~\ref{proposition:Completion}, and
		proceed like in the first case.
	For (6.b),
		first suppose that $H\preccurlyeq H'$.
	Thus $H'=H\curlyvee(H'\curlywedge\ocompl{H})$ by orthomodularity.
	Hence:
		$$\begin{array}{@{}r@{\ \ }c@{\ \ }l@{\qquad}l@{}}
			\rho(H')	&=& \rho(H\curlyvee(H'\curlywedge\ocompl{H})) & \\[2\jot]
						&=& \rho\left(\ocompl{\left(\ocompl{H}\curlywedge\ocompl{\left(H'\curlywedge\ocompl{H}\right)}\right)}\right) & \text{(definition)}\\[4\jot]
						&=& \acompl\rho\left(\ocompl{H}\curlywedge\ocompl{\left(H'\curlywedge\ocompl{H}\right)}\right) & \text{($\acompl$-homomorphism)}\\[3\jot]
						&=& \acompl\left(\rho(\ocompl{H})\cap\rho\left(\ocompl{\left(H'\curlywedge\ocompl{H}\right)}\right)\right) & \text{(meet homomorphism)}\\[3\jot]
						&=& \acompl\rho(\ocompl{H})\cup\acompl\rho\left(\ocompl{\left(H'\curlywedge\ocompl{H}\right)}\right) & \text{(2)}\\[2\jot]
						&=& \acompl\acompl\rho(H)\cup\acompl\acompl\rho(H'\curlywedge\ocompl{H}) & \text{($\acompl$-homomorphism)}\\[\jot]
						&=& \rho(H)\cup\rho(H'\curlywedge\ocompl{H}) & \text{(1)}\\[\jot]
						&=& \rho(H)\cup(\rho(H')\cap\rho(\ocompl{H})) & \text{(meet-homomorphism)}\\[\jot]
						&=& \rho(H)\cup(\rho(H')\cap\acompl\rho(H)) & \text{($\acompl$-homomorphism)}
		\end{array}$$
	Now suppose for (6.b) that $\rho(H)\subseteq\rho(H')$.
	Hence $H\preccurlyeq H'$ by Proposition~\ref{proposition:Completion}, and
		proceed like in the first case.
	For (7), consider (4).
	Hence: 
		$$\begin{array}{@{}r@{\ \ }c@{\ \ }l@{\qquad}l@{}}
			\acompl\rho(0)	&=& \acompl(\rho(H)\cap\acompl\rho(H)) & \text{}\\[\jot]
							&=& \acompl\rho(H)\cup\acompl\acompl\rho(H) & \text{(2)}\\[\jot]
							&=& \acompl\rho(H)\cup\rho(H) & \text{(1)}\\[\jot]
							&=& \rho(1) & \text{(5)}
		\end{array}$$
	For (8), consider (7).
	Thus $\acompl\acompl\rho(0)=\acompl\rho(1)$.
	Hence  $\rho(0)=\acompl\rho(1)$ by (1).
\end{proof}
\noindent
In spite of Proposition~\ref{proposition:PropertiesOfSim} having only the status of a proposition, 	
	its proof actually contains more information than 
		the proof of the following (main) theorem.
\begin{theorem}[Representation Theorem for Orthomodular Lattices]\label{theorem:Representation}
	The structure 
		$\mathfrak{S}:=\langle\,\{\,\rho(H) \mid H\in\mathrm{H}(\states)\,\},\rho(0),\cap, \cup,\rho(1),\acompl,\subseteq\,\rangle$ is a \emph{sublattice of sets} of the powerset lattice $\BAO$ that 
			is \emph{isomorphic} to $\OML$ via 
			the \emph{\textbf{lattice}-embedding} $\rho$, that is, 
				$\rho$ is a bijection between $\OML$ and $\mathfrak{S}$, and preserves the structure: 
			$$\begin{array}{@{}r@{\ \ }c@{\ \ }l@{}}
			\rho(0) &=& \rho(0) \\[\jot]
			\rho(H\curlywedge H') &=& \rho(H)\cap\rho(H') \\[\jot]
			\rho(H\curlyvee H')	&=& \rho(H)\cup\rho(H') \\[\jot]
			\rho(1) &=& \rho(1) \\[\jot]
			\rho(\ocompl{H}) &=& \acompl{\rho(H)}
		\end{array}$$
\end{theorem}
\begin{proof}
	By definition, 
		$\rho$ is an injection from $\OML$ into $\BAO$ and thus also into $\mathfrak{S}$, and
		preserves the orthocomplement $\ocompl{}$ and the orthomodular meet $\curlywedge$.
	By definition of $\mathfrak{S}$, 
		$\rho$ is also a surjection from $\OML$ onto $\mathfrak{S}$, and
		preserves also the orthomodular bounds $0$ and $1$.
	For the preservation of the orthomodular join $\curlyvee$ consider that:
	$$\begin{array}{@{}r@{\ \ }c@{\ \ }l@{\qquad}l@{}}
		\rho(H\curlyvee H') &=& 
			\rho\left(\ocompl{\left(\ocompl{H}\curlywedge\ocompl{H'}\right)}\right) & \text{(definition)}\\[\jot]
			&=& \acompl\rho(\ocompl{H}\curlywedge\ocompl{H'}) & \text{($\acompl$-homomorphism)}\\[\jot]
			&=& \acompl(\rho(\ocompl{H})\cap\rho(\ocompl{H'})) & \text{(meet homomorphism)}\\[\jot]
			&=& \acompl\rho(\ocompl{H})\cup\acompl\rho(\ocompl{H'}) & \text{(Proposition~\ref{proposition:PropertiesOfSim}.2)}\\[\jot]
			&=& \acompl\acompl\rho(H)\cup\acompl\acompl\rho(H') & \text{($\acompl$-homomorphism)}\\[\jot]					&=& \rho(H)\cup\rho(H') & \text{(Proposition~\ref{proposition:PropertiesOfSim}.1)}	
		\end{array}$$	
\end{proof}
\noindent
Let us take stock, and record 
	which properties of the accessibility relation $R$ were actually required to prove our theorem.
Observe that only 
	Proposition~\ref{proposition:PropertiesOfSim}.1 and 
	\ref{proposition:PropertiesOfSim}.2 require such properties.
The proof of Proposition~\ref{proposition:PropertiesOfSim}.2 requires 
	Fact~\ref{fact:PropertiesOfPosOp} and only that one as such a property.
Less obviously,
	because somehow hidden in plain sight, 
		Proposition~\ref{proposition:PropertiesOfSim}.1 
			\emph{itself} is actually another such property.
It stipulates that 
	for all $S\in\{\,\rho(H) \mid H\in\mathrm{H}(\states)\,\},$
	$$\begin{array}{@{}r@{\ \ }c@{\ \ }l@{}}
				S &=&	(\posop{R}\circ\compl{\,\cdot\,}\circ\posop{R}\circ\compl{\,\cdot\,})(S)\\[\jot]
				  &=&	(\posop{R}\circ\necop{R}\circ\compl{\,\cdot\,}\circ\compl{\,\cdot\,})(S)\\[\jot]
				  &=&	(\posop{R}\circ\necop{R})(S)\,.
		\end{array}$$
It is well known that 
	the inclusion $(\posop{R}\circ\necop{R})(S)\subseteq S$ stipulates the \emph{symmetry} of $R$, 
		corresponding to the so-called \emph{B-axiom} $A\rightarrow\Box\Diamond A$ 
			(see for example \cite{ModalProofTheory}).
However to our knowledge, 
	the inclusion $S\subseteq(\posop{R}\circ\necop{R})(S)$ has no name yet;
		we shall call it the \emph{\textbf{Q-property}} of $R$, 
		its corresponding axiom $\Box\Diamond A\rightarrow A$ the \emph{\textbf{Q-axiom,}} and 
			the resulting normal modal logic the logic \emph{\textbf{BQ}} (= K+B+Q).
Observe that
	both the B-axiom as well as the Q-axiom are Sahlqvist-formulas \cite{Sahlqvist}, and 
		so correspond to (satisfiable) first-order-logically (FOL-)definable classes of Kripke-frames.
Of course,
	symmetry is FOL-definable: 
		$\forall s\forall s'(s\mathrel{R}s' \rightarrow s'\mathrel{R}s)$.
And so is the class of Kripke-frames corresponding to the Q-axiom \cite{SQEMA}: 
	$\forall s\exists s'(s\mathrel{R}s'\land\forall s''((s'\mathrel{R}s'')\rightarrow(s''=s)))$, which
		implies the seriality $\forall s\exists s'(s\mathrel{R}s')$ of $R$ 
			(see also Appendix~\ref{appendix:seriality}), which in turn 
				corresponds to the D-axiom $\Box A\rightarrow\Diamond A$ (or equivalently, $\neg\Box\bot$).
That is,
	falsehood can never be observed to be true (consistency of observability).
Note that the satisfiability of these FOL-translations implies 
	the existence of our accessibility relation $R$, which
		is characterised in terms of these translations.

In the light of \cite{QLnonelementary},
	the FOL-definability of our accessibility relation $R$ may seem contradictory.
However,
	consider that \cite{QLnonelementary} applies to 
		Kripke-semantics of QL and not to
			Kripke-semantics of some modal logic (such as BQ) into which QL embeds.
The properties of the two accessibility relations need not coincide.
For a well-known counter-example consider 
	Intuitionistic Logic (IL), whose
		Kripke-semantics has an accessibility relation that is a \emph{partial} order \cite{KripkeSemanticsIL}, and
	the normal modal logic S4, into which IL embeds via the G\"odel-McKinsey-Tarski translation, and
		whose accessibility relation is only a \emph{pre-}order.

In the following corollary,
	we apply $\rho$ tacitly on QL-formulas $A$ rather than 
		on their semantic counterparts $H$ (their Lindenbaum-Tarski algebra quotient).
\begin{corollary}\label{corollary:reduction}
	Syntactically, 
		$\rho$ is a \emph{linear-time reduction} from QL to BQ: 
			$$\text{$A\in\mathrm{QL}$ if and only if $\rho(A)\in\mathrm{BQ}$.}$$
\end{corollary}
\begin{proof}
	Assuming that any quantum disjunction in a given QL-formula $A$ has been expanded by 
	its definition in terms of quantum negation and quantum conjunction, 
we just substitute 
	any occurrence of the quantum-negation symbol $\acompl$ in 
	$A$ with $\neg\Box$ or with $\Diamond\neg$ (as already said in the introduction), in 
		order to obtain the corresponding BQ-formula.
When performed on $A$ represented as a (linear) string of symbols, 
	this substitution procedure obviously takes linear time (no back-tracking required).
\end{proof}

In the sequel, let us write the customary ``$\dedBQ A$'' for ``$A\in\mathrm{BQ}$,''
	meaning that $A$ is a theorem of BQ.
The following proposition concludes that $\Box$ and $\Diamond$ are globally equivalent.
\begin{proposition}\ 
	\begin{enumerate}
		\item The converse of necessitation
			\begin{center}
				from $\Box A$ infer $A$\qquad(CN), 
			\end{center}
				is a derived deduction rule for BQ.
		\item $\text{$\dedBQ\Box A$ iff $\dedBQ A$}$
		\item $\text{$\dedBQ A$ iff $\dedBQ\Diamond A$}$
		\item $\text{$\dedBQ\Box A$ iff $\dedBQ\Diamond A$}$
	\end{enumerate}
\end{proposition}
\begin{proof}
	For (1), suppose that $\dedBQ\Box A$.
	Hence $\dedBQ\Diamond A$, by \emph{modus ponens}, because $\dedBQ\Box A\rightarrow\Diamond A$.
	Hence $\dedBQ\Box\Diamond A$, by necessitation.
	Hence $\dedBQ A$, by \emph{modus ponens}, because $\dedBQ\Box\Diamond A\rightarrow A$ (Q-axiom).
	That is,
		$\dedBQ\Box A$ implies $\dedBQ A$.
	From this and necessitation then follows (2).
	For (3) and (4), consider:
		$$\begin{array}{rcl@{\qquad}l}
			\dedBQ\Diamond A & \text{iff} & \dedBQ\Box\Diamond A & \text{(2)}\\[\jot]
							 & \text{iff} & \dedBQ A & \text{BQ-axiom, \emph{modus ponens}, bidirectional}\\[\jot]
							 & \text{iff} & \dedBQ \Box A & \text{(2)}
		\end{array}$$
	
\end{proof}

\section{Conclusion}
We have demonstrated that 
	Quantum Logic (QL) is a fragment of the classical normal modal logic BQ, which 
		in turn is a fragment of classical (in every sense of the word) first-order logic 
			(via the Standard Translation \cite{ModalLogicSemanticPerspective}).
	In other words, 
		we have refuted Birkhoff and von Neumann's classic thesis that 
			the logic (the formal character) of Quantum Mechanics would be non-classical \cite{BirkhoffVonNeumann} as well as 
		Putnam's thesis that quantum logic (of his kind) would be the correct logic for propositional inference in general \cite{PutnamQL}.
	The propositional logic of Quantum Mechanics has turned out to be modal but classical, and
		the correct logic for propositional inference need not have an extroverted quantum character.
The philosophical key to our result has been to 
	\emph{internalise observability} into our (logical) system (by means of a normal necessity modality), 
		which in some sense is what Quantum Mechanics has always told us to do.
With that,
	the mystery of the failure of classical conjunction to distribute over classical disjunction
		has dissolved and 
	an elementary-logical solution for this (weak)\footnote{%
		As opposed to a strong paradox (a formal contradiction/inconsistency),
			a weak paradox dissolves upon (proper) formalisation.} paradox has emerged.
(Other paradoxes of Quantum Mechanics may subsequently dissolve too.)
In a formal sense,
	we have 
		reduced Quantum Logic (QL) to Classical Logic, 
			within a simple modal logic (BQ).
Translations of QL in a similar spirit but to more complex modal systems can be found in \cite{QLintoBr} and \cite{DynamicQL}.
Then, translations of ortholattices 
	(lattices with an orthocomplement but, 
		as opposed to our orthomodular lattices, 
			without the orthomodular law) can be found in  
				\cite{OrthologicSemantically} and \cite{QuantumLogics}.
Finally, first-order extensions (with quantifiers) as well as 
	dynamic and fixpoint extensions are possible 
		(see for example \cite{FOModalLogic}, \cite{DynamicQL}, and \cite{ModalMuCalculi}, respectively).

\paragraph{Acknowledgements}
I thank Norman Megill and  Mladen Pavi\v{c}i\'{c} for 
	pointing out an erroneous Corollary~\ref{corollary:reduction} in  
			the first arXiv-version of this paper.
I take the sole responsibility for this error.
Then,  
	I also thank 
		Robert Goldblatt, Denis Saveliev, and Ronnie Hermens for discussions that
			have lead to improvements of the second arXiv-version of this paper.

\bibliographystyle{plain}

\appendix

\section{Proof that seriality is a theorem of BQ}\label{appendix:seriality}
The following lemma recalls that 
	so-called \emph{regularity} is a derived rule for  
		any normal modal logic and thus also for BQ.
\begin{lemma}[Regularity]
	The rule 
		``from $A\rightarrow B$ infer $\Box A\rightarrow\Box B$,'' called \emph{regularity,}
			is derivable in any normal modal logic.
\end{lemma}
\begin{proof}
	Suppose that $A\rightarrow B$ is a theorem of the considered normal modal logic, 
		say $\mathrm{L}$, that is, $(A\rightarrow B)\in\mathrm{L}$.
	Hence, $\Box(A\rightarrow B)\in\mathrm{L}$, 
		by the necessitation rule for $\mathrm{L}$.
	Of course,
		$(\Box(A\rightarrow B)\rightarrow(\Box A\rightarrow\Box B))\in\mathrm{L}$, 
			by the Kripke-axiom of $\mathrm{L}$.
	Hence, 
		$(\Box A\rightarrow\Box B)\in\mathrm{L}$, 
			by the rule of \emph{modus ponens} for $\mathrm{L}$.
\end{proof}
\noindent
Now, seriality can be proved in BQ in the following five lines:
\begin{enumerate}
	\item $\Box\Diamond\bot\rightarrow\bot$\hfill instance of the Q-axiom\\[-6\jot]
	\item $\neg\Box\Diamond\bot$\hfill 1, \emph{modus tollens}\\[-6\jot]
	\item $\bot\rightarrow\Diamond\bot$\hfill ``falsehood implies everything''\\[-6\jot]
	\item $\Box\bot\rightarrow\Box\Diamond\bot$\hfill 3, regularity\\[-6\jot]
	\item $\neg\Box\bot$\hfill 2, 4, \emph{modus tollens}
\end{enumerate}

\end{document}